# In Silico Pharmacokinetic and Molecular Docking Studies of Natural Plants against Essential Protein KRAS for Treatment of Pancreatic Cancer

**Marsha Mariya Kappan\* and Joby George**

Department of Computer Science and Engineering, Mar Athanasius College of Engineering, Kothamangalam, Ernakulam - 686666, Kerala, India; kappanmarsha@gmail.com

## Abstract

A kind of pancreatic cancer called Pancreatic Ductal Adenocarcinoma (PDAC) is anticipated to be one of the main causes of mortality during past years. Evidence from several researches supported the concept that the oncogenic KRAS (Ki-ras2 Kirsten rat sarcoma viral oncogene) mutation is the major cause of pancreatic cancer. KRAS acts as an on-off switch that promotes cell growth. But when the KRAS gene is mutated, it will be in one position, allowing the cell growth uncontrollably. This uncontrollable multiplication of cells causes cancer growth. Therefore, KRAS was selected as the target protein in the study. Fifty plant-derived compounds are selected for the study. To determine whether the examined drugs could bind to the KRAS complex's binding pocket, molecular docking was performed. Computational analyses were used to assess the possible ability of tested substances to pass the Blood Brain Barrier (BBB). To predict the bioactivity of ligands a machine learning model was created. Five machine learning models were created and have chosen the best one among them for analyzing the bioactivity of each ligand. From the fifty plant-derived compounds the compounds with the least binding energies are selected. Then bioactivity of these six compounds is analyzed using Random Forest Regression model. Adsorption, Distribution, Metabolism, Excretion (ADME) properties of compounds are analyzed. The results showed that borneol has powerful effects and acts as a promising agent for the treatment of pancreatic cancer. This suggests that borneol found in plants like mint, ginger, rosemary, etc., is a successful compound for the treatment of pancreatic cancer.

## 1. Introduction

Within the next ten years, Pancreatic Ductal Adenocarcinoma (PDAC) is expected to become the second most prevalent cause of mortality. Despite advances in diagnostics and novel chemotherapy treatments, the prognosis for this disease is bad. The main factor causing pancreatic cancer is the mutation of the oncogenic KRAS. It permanently activates the KRAS protein, which in pancreatic cancer acts as a tumour suppressor. The mutation of KRAS may cause KRAS activation and this in turn acts as a molecular switch to induce cell proliferation, transformation, etc. Pancreas is deep inside our body. So, it is hard to find any growths or lumps there. A highly malignant tumour of the digestive tract, Pancreatic Cancer (PC) presents substantial challenges for early detection and therapy[1,2]. Patients with PC have the fourth highest death rate in the United States, according to statistics from the American Cancer Society[3-5]. According to estimates, roughly 57,600 people will be diagnosed with PC in 2020, and approximately 47,050 people will die from the condition, making PC an incurable disease. Regardless of the PDAC stage, KRAS mutations are present in a higher number of patients and are associated with a worse overall survival rate. When bound to Guanosine Triphosphate (GTP) KRAS is active, and when it is bound to Guanosine Diphosphate (GDP), it is deactivated. GTP nucleotide is hydrolysed by GTPases, which are hydrolase enzymes, into Guanosine Diphosphate (GDP). KRAS is a member of the RAS family. The RAS family includes two related genes called HRAS and NRAS whose mutation also causes human cancer. These mutations severely reduce the Ras GTPase's ability to function, thereby locking the Ras protein in an active, GTP-bound state. Over 10% of all human cancers are caused by mutations in the three Ras genes taken together. A major reason for RAS mutation is because RAS signals to several effector pathways whose activation promotes oncogenic transformation. This includes the small GTPases Rho, Rac, and Ral, as well as phospholipase C and the MAP-Kinase (MAPK) and PI3K/AKT/mTOR pathways. These pathways together will promote cell growth, cell survival, metabolism, etc. By continuously activating these pathways, the RAS oncogene gives the cancer cell advantages for growth and survival. A variety of biological samples, such as fresh and fixed tumour tissue or biopsy samples, fine-needle aspiration materials, cytological samples, blood, and plasma, can contain KRAS mutations[6,7]. Many synthetic medicines are available for the treatment of pancreatic cancer. Medications and Chemotherapy can be used to suppress the growth and division of tumour cells. Different Food and Drug Administration (FDA) approved drugs such as Capecitabine, Fluorouracil, Gemcitabine, etc. are some drugs used in chemotherapy. These drugs are given one at a time or as a combination of more than one. There are numerous adverse effects associated with these chemotherapy medications, including nausea, gastrointestinal issues, rash, fatigue, mouth sores, and hair loss. This can also cause low white blood cell, red blood cell, and platelet counts which result in increasing the risk of anemia, infections, and easy bleeding. Synthetic drugs have more side effects and reactions. Paracetamol is a drug used in the treatment of fever and to relieve pain, but it can also cause liver poisoning as a side effect[8]. The development of synthetic drugs undergoes several methodologies in laboratories as they are not found in nature. Natural compounds can be used as drugs. It has fewer side effects compared to synthetic medicines. Compounds found in plants such as turmeric, cloves, and cinnamon have healing effects. For e.g., curcumin in turmeric has healing effects for inflammation, pain, arthritis, etc. Treatments using natural compounds are safer. But proper preparations and clinical trials are required, otherwise, poisoning can also happen[9]. The use of machine learning and bioinformatics can make drug discovery fast and helps to use natural products to obtain suitable drugs. Drug discovery is a long-term costly process. The chance of failure in drug discovery is high as most drugs fail while entering clinical trials. Computational studies in drug discovery will help to overcome this problem. Such techniques will help to reduce the number of ligands needed to be screened in experiments. It also helps to reduce the cost by ensuring that the best ligand only enters clinical studies. These computer-aided tools can also predict toxicity and side effects. The Quantitative Structure-Activity Relationship (QSAR) is a method that uses statistical or machine learning methods to predict biological activities for compounds of interest as a function of their descriptors[10]. Molecular docking can be used to identify the best-fit orientation of ligands to the target protein. Molecular docking helps to identify the binding energy also. To reveal the KRAS protein's inhibitory activity, QSAR models were developed using Machine Learning (ML) techniques (such as random forest), and descriptors (such as molecular fingerprints) are calculated. Different ML models are created and among them best one is chosen for QSAR analysis. A varied group of plant-derived compounds is chosen from natural plants to find active KRAS inhibitors. The ligand-based and structure-based approaches used in this study are expected to aid in the discovery and development of effective KRAS inhibitors.

---

*\*Author for correspondence*

There are some studies that are focused on drug discovery for the treatment of pancreatic cancer. In their research, Pierre Cordelier, Barbara Bournet, and Louis Buscail examined the causes, symptoms, and prognosis of pancreatic cancer[11]. Studies on patients with PDAC have focused on the detection of circulating tumour cells, cell-free ctDNA, exosomes, and tumour-induced platelets. The KRAS mutation is the primary target for ctDNA detection and DNA carried into exosomes. Mutations of KRAS in plasma and serum can be detected for monitoring cancer. This makes it unsurprising that KRAS mutation surveillance in blood from PDAC patients is gaining momentum[12,13] if the KRAS mutation's presence or absence can affect the prognosis of PDAC[14-18]. Overall, whether or not patients with PDAC receive surgery (with total tumour excision or locally progressed and/or metastatic PDAC), the presence of the mutated KRAS has an impact on their prognosis[14-16,18]. It should be noted that various studies' health records vary and may contain individuals with recovered or treated PDAC and ampullary carcinomas, nonresectable PDAC, and a previously removed tumour that later reappeared. Roughly mutations in KRAS cause one-third of most cancers. KRAS controls cell growth by cycling between on and off states when needed, but G12C mutation turns it into a state which promotes uncontrollable cell growth. So, the aim was to find a drug that could attach to KRAS G12C protein and turn it off. Sotorasib is a KRAS inhibitor approved by the FDA for the treatment of people with lung cancer that has KRAS G12C mutation. Several other KRAS G12C inhibitors are undergoing clinical trials[19]. Isolated phytochemicals from various plant extracts can be employed as chemotherapeutics since they have a range of anti-tumour, anti-inflammatory, anti-oxidant, and anti-bacterial activities[20]. Numerous phytochemicals, including vinca alkaloids, taxanes, *Camptothecin* derivatives, *Cephalotaxus, Colchicine, Ellipticine, Berberine, Combretastatins,* and triterpenoid acids, demonstrated potential against cancer. Systems pharmacology is one of the most well-known new methods for researching how medications interact with biological systems. Systems pharmacology has recently been used to pinpoint diverse natural compounds as well as their mechanisms[21]. Quantitative Systems Pharmacology (QSP) links systems biology with Pharmacokinetics (PK) and Pharmacodynamics (PD) to fully understand a drug's effectiveness and toxicity in complicated disease systems like cancer.

## 2. Materials and Methods

Figure 3 depicts a flowchart of the study's workflow. In a nutshell, this contains a broad QSAR model for predicting and evaluating KRAS inhibition. This includes a data set with a defined endpoint, a clear learning methodology, a QSAR model, and a deterministic interpretation of the QSAR model. A chemically varied data set was used to perform molecular docking in order to comprehend the actual binding mechanism. Molecular docking is one of the important steps and it is done for all the ligands. That is all the ligands are docked with the target protein KRAS and their binding energies are evaluated. The work has two parts: part 1 consists of model creation. Users can check how a particular molecule act in KRAS inhibition using the created model. Users can insert SMILES notation of a particular molecule as input. Then they will get the bioactivity data of that molecule. Part 2 consists of molecular docking and drug likeliness prediction and detection of ADME properties.

### 2.1 Finding Out the Target Protein

The first step is to identify the target protein. The KEGG pathway is used to find the target protein. The KEGG is a database that contains information about genomes, biological pathways, diseases, medications, and chemical compounds. So, we stated the name of the disease, which is pancreatic cancer, in the KEGG pathway. The disease's KEGG pathway is then displayed. Different target proteins were discovered as a result of this. And it was from there that the KRAS protein was chosen for further research. From the KEGG pathway, we have identified target proteins for pancreatic cancer. They are K-Ras, Her2/neu, Smad4, BRCA2, p16, p53. From these, we have selected the KRAS[11] oncogene for our studies since there are only limited studies have taken place in the case of KRAS protein. And also, the KRAS protein may cause many other different cancers also. The primary cause of pancreatic cancer is an oncogenic KRAS mutation. This mutation causes the KRAS protein to become permanently activated, acting as a molecular switch to switch on several intracellular signaling pathways and transcriptional factors that support cell migration, proliferation, transformation, and also survival.

### 2.2 Ligand Preparation

The compounds found in different plants are selected as ligands for studies. Fifty ligands from different plant sources are selected for the study. Then the structure of these ligands is drawn using ChemSketch software. And the file is saved using .mol format. Then this .mol file is converted to the .pdbqt format. The particular file is used for molecular docking. The ligands used and from which plant source it is extracted are given in Table 1.

**Table 1.** Ligands and their sources

| Sl no. | Ligand name | Binding Energy | Ligand Source |
|---|---|---|---|
| 1 | Alpa spinasterol | +264.47 | *Bupleurum falcatum* |
| 2 | Ginsenosides | +409.20 | *Panax ginseng* |
| 3 | Codeine | +159.24 | *Papaver somniferum* |
| 4 | Isoliquiritin | +407.41 | *Glycyrrhiza glabra* (Licorice) |
| 5 | Glabridin | +215.12 | *Glycyrrhiza glabra* |
| 6 | Liquiritin | +318.17 | *Glycyrrhiza glabra* |
| 7 | Gingerol-6 | +51.58 | *Zingiber officinale* (Ginger) |
| 8 | Gingerol-4 | +37.63 | *Zingiber officinale* |
| 9 | Shogaol | +54.78 | *Zingiber officinale* |
| 10 | Esculetin | +4.53 | *Bupleurum falcatum* |
| 11 | Beta sitosterol | +316.71 | *Panax ginseng* |
| 12 | Geraniol | +5.83 | *Zingiber officinale* |
| 13 | Farnesal | +42.98 | *Zingiber officinale* |
| 14 | Beta citronellol | +1.71 | *Zingiber officinale* |
| 15 | Guaiacol | -1.34 | *Guaiacum officinale* |
| 16 | Hispaglabridin A | +707.92 | *Glycyrrhiza glabra* |
| 17 | Chrysin | +131.06 | *Phyllanthus niruri* |
| 18 | Thannilignan | +83.83 | *Phyllanthus niruri* |
| 19 | Scutellarein | +122.76 | *Phyllanthus niruri* |
| 20 | Naringin | +152.02 | *Vitis vinifera* (Grapefruit) |
| 21 | Hispaglabridin B | +730.19 | *Glycyrrhiza glabra* |
| 22 | Glycyrrhizin | +2555.86 | *Glycyrrhiza glabra* |
| 23 | Kuwanon S | +263.04 | *Phyllanthus niruri* |
| 24 | Ellagic acid | +195.35 | *Phyllanthus niruri* |
| 25 | Anolignan | +117.13 | *Phyllanthus niruri* |
| 26 | Belotecan | +472.89 | *Camptothecin* analogue |
| 27 | Borneol | +3.51 | *Zingiber officinale* |
| 28 | Apigenin | +83.47 | *Lycopodium clavatum* |
| 29 | Coriandrin | +61.90 | *Coriandrum sativum* (Coriander) |
| 30 | Curcumin | +87.58 | *Curcuma longa* (Turmeric) |
| 31 | Glabridin | +160.43 | *Glycyrrhiza glabra* |
| 32 | Liquiritigenin | +74.24 | *Glycyrrhiza glabra* |
| 33 | Quercetin | +103.98 | *Camellia sinensis* (Green tea) |
| 34 | Ursolic acid | +501.31 | *Ocimum basilicum* (Basil) |
| 35 | Topotecan | +860.78 | *Camptothecin* analogue |
| 36 | Quinic acid | +0.70 | *Artocarpus heterophyllus* (Jackfruit) |
| 37 | Luteolin | +91.50 | *Ocimum tenuiflorum* (Tulasi) |
| 38 | Apigenin | +83.04 | *Ocimum tenuiflorum* |
| 39 | Rosmarinic acid | +74.54 | *Ocimum tenuiflorum* |
| 40 | Nimbolide | +449.82 | *Ocimum tenuiflorum* |
| 41 | Gallicacid | +2.63 | *Ocimum tenuiflorum* |

**Table 1. To be continued...**

| Sl no. | Ligand name | Binding Energy | Ligand Source |
|---|---|---|---|
| 42 | Diterpene | +76.86 | *Coleus forskohlii* |
| 43 | Epicatechin | +103.87 | *Azadirachta indica* (Neem) |
| 44 | Gedunin | +729.51 | *Azadirachta indica* |
| 45 | Nimbolide | +488.76 | *Azadirachta indica* |
| 46 | Podophyllotoxin | +478.76 | *Podophyllum peltatum* (Mayapple) |
| 47 | Epigallocatechin | +278.09 | *Vitis vinifera* (grapes), |
| 48 | Irinotecan | +1817.61 | *Camptothecin* analog |
| 49 | Oleic acid | +62.35 | *Helianthus* (Sunflower) |
| 50 | Germacrone | +65.19 | *Linnaeus* (Koova) |

### 2.3 Molecular Docking

The structure of KRAS (PDB ID: 6MNX) was selected from the Protein data bank and then it is modified initially by eliminating the water molecules and possible side chains. The docking is performed using AutoDock software. By re-establishing bonds and adding missing hydrogen atoms, the protein was created. One receptor chain has been chosen by us. The target protein was then purified by combining the atomic charge, eliminating non-polar hydrogen atoms, non-standard amino acid residues, and lone pair atoms. Then check for missing atoms and if found repair the missing atoms. AutoDock tool is used to add data on essential hydrogen atoms, add Kollman charges, and solvation parameters.

The grid specifies the region where the ligand dock with protein. Here we don't know the binding site so we go for blind docking. In blind docking, the grid box is set up to cover the entire protein. We are considering only one chain, chain A. The grid box is configured to cover the KRAS protein with a dimension of (X, Y, and Z axes of 93.432, 21.305, and 28.993 respectively). The AutoDock tool was used to accomplish molecular docking. Here docking is performed with 50 plant-derived compounds. Each of these ligands is docked with the target protein and the binding energies are analysed. Lamarckian Genetic Algorithm (LGA) is used to generate docking simulations. The number of GA runs is set as 50 and the population size is set as 300 for better results of the docking experiment. After successful docking, many files will be created in the docking folder. Then the created .dlg file is used to analyse binding energies.

Perform molecular docking for all the ligands and from each of the generated DLG files select the row with the least binding energy. So finally, we will get 50 minimum binding energies since we have 50 ligands. Table 1 consists of the binding energies of each ligand after docking. From Table 1 select the ligands with the least binding energy. So, from these 50 binding energies sort out 6 rows that have the least binding energies. Then we can identify these 6 ligands with the least binding energies are the best. When the energy is more negative, the ligand will be better. The lesser the binding energy, the better will be the binding of the ligand and protein. We must find out the ligands with minimum binding energy. So, we have selected six ligands which have minimum binding energy. The ligands selected for study and its source are given in Table 1. The structure of ligands is given in Figure 1. From Table 1 the ligands which have the least binding energy are quinic acid, beta citronellol, gallic acid, borneol, esculetin, and geraniol. After performing molecular docking, we visualize binding pockets. Open the .dlg file of a particular ligand in AutoDock software. Then open the macromolecule file. Then click on play. From the RMSD table in the .dlg file select the row with minimum binding energy. Look its run. Then adjust this run in Autodock and then write the structure. The structure will be in .pdb file format. Using OpenBabel software first convert this .pdb file to .pdbqt file. This is done because PLIP will take only files in. pdb format. When "Analyse" on Protein-Ligand Interaction Profiler (PLIP) was selected, a three-dimensional rendering of the protein-ligand docked structure appeared. PLIP is an automated tool designed for high-throughput investigation and visualisation of pertinent non-covalent interactions in three-dimensional structures. When a protein and ligand combination is submitted, PLIP produces several distinct interactions between the protein and ligand that stabilises the system. After performing molecular docking, we visualize binding pockets. The results obtained from Autodock software are uploaded to PLIP (https://plip-tool.biotec.tu-dresden.de/plip-web/plip/index). The structure of the ligand is in .pdb file format. Using OpenBabel software first convert this .pdb file to .pdbqt file. This is done because PLIP will take only files in. pdb format. Then using PLIP a three-dimensional rendering of the protein-ligand docked structure appeared. When a protein and ligand combination is submitted, PLIP produces several distinct interactions between the protein and ligand that stabilise the system.

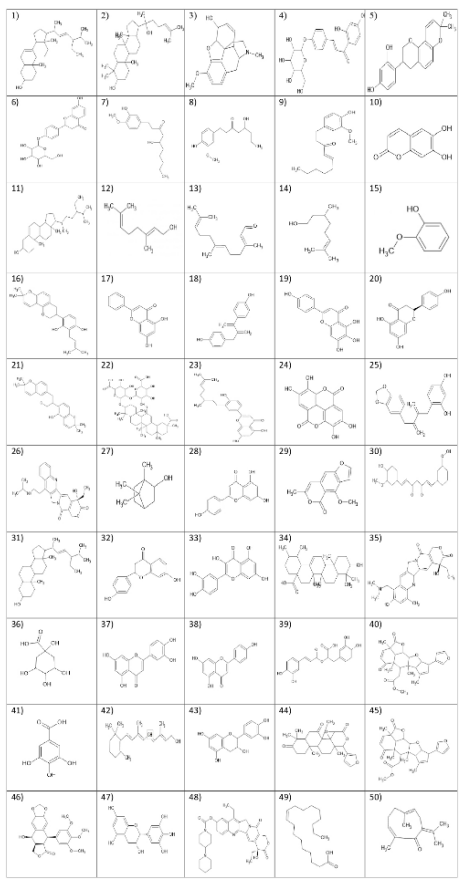

**Figure 1.** Structure of selected fifty plant derived compounds.

Figure 2 represents images of selected six ligands when docked with target protein KRAS, generated using PLIP. The ligands used were a) quinic acid (CHEMBL465398), b) beta citronellol (CHEMBL395827), c) gallic acid (CHEMBL288114), d) borneol (CHEMBL486208), e) esculetin (CHEMBL244743), and f) geraniol (CHEMBL25719).

### 2.4 Dataset Preparation

This is the machine learning model creation part. The dataset is selected from the ChEMBL database. A dataset of inhibitors against the KRAS was taken from the ChEMBL database. For that, we need to install the ChEMBL web resource client package in the notebook (Jupyter Notebook) environment so that we can retrieve bioactivity data from the ChEMBL database for creating our machine learning models. Jupyter Notebook is used for machine learning model creation. The targets that we obtain from the ChEMBL datreferrefer to the target protein or target organism that the drug will act on. After searching we got three targets. We have selected any one target with the target type as Single Protein. We have selected the target with ChEMBL ID (CHEMBL2189121) and name:

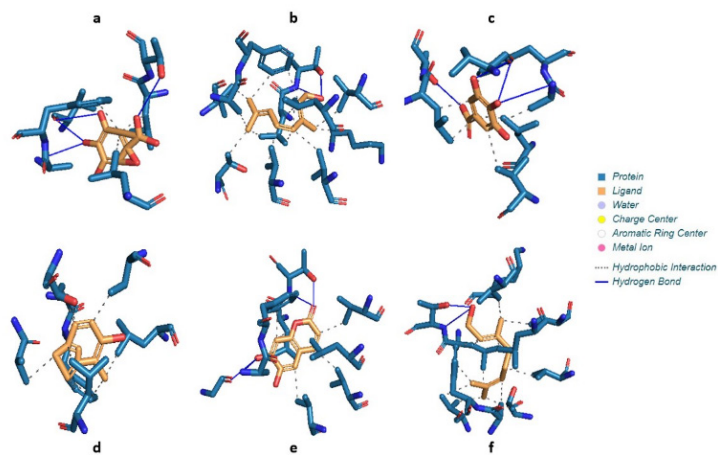

**Figure 2.** Selected six ligands after docking with target protein KRAS, created using PLIP.

GTPase KRas. The dataset contains many compounds and a compound means a drug or a molecule which produces a modulatory activity i.e. it affects the target protein in some way to produce a desired biological effect which cures the symptoms. Here we will retrieve bioactivity data for only GTPase KRAS (CHEMBL2189121) based on the IC50 values represented in the nanomolar (nm) unit. The standard value of standard type (IC50) represents the potency of the drug. The lower the value better is the potency. Inhibitory concentration at 50% will have a low concentration. That means to obtain 50% of inhibition of a target we need a lower concentration of the drug. If the IC50 value is high then we need more amount of drugs to produce the same inhibition at 50%. Next, take care of the missing data. Delete a specific row if any compounds in the standard value column have missing values. Then categorize the data as part of pre-processing. So, classify compounds into three bioactivity classes such as active, inactive or intermediate.

Compounds that have an IC50 value of less than 1000nm are classified as active and compounds with IC50 value greater than 10000nm are classified as inactive and those that have values between 1000nm and 10000 nm are classified as intermediate. If more than one compound with the same molecular ChEMBL id is present that is also removed to reduce redundancy in the dataset. After removing the redundancies and missing data combine the ChEMBL id, SMILES notation, bioactivity class and standard value into a data frame. Then save the data frame to the CSV file. This CSV file is used for building machine-learning models. The goal of the drug discovery is to find a compound or molecule that will be able to inhibit the function of the KRAS protein. The prepared CSV file is used for creating all five machine learning models.

## 2.5 QSAR Modelling

QSAR is a technique that applies machine learning to learn the relationship between chemical structure and biological activity. Select many molecules (100 or 1000 or more). Each molecule has a chemical structure from which we can calculate its molecular fingerprints. We use PubChem fingerprint which is represented using binary 0's and 1's. In the data frame the X descriptors correspond to the molecular descriptors and the Y variable corresponds to the biological activity. 1 means the molecule is active and 0 means inactive. This dataset is used to train a machine learning model. The machine learning model can learn the chemical structure and the biological activity. So later when we give a molecule with a given molecular descriptor as input to the model the model will predict whether the molecule is active or inactive. The model will also be able to provide insights into which features are important. QSAR modelling includes different steps such as Exploratory Data Analysis (EDA), Mann-Whitney U test, Fingerprint Calculation and Model Building.

## 3. Results

### 3.1 Exploratory Data Analysis

Here we apply data science for drug discovery. Here we need to compute molecular descriptors and then perform Exploratory Data Analysis (EDA) on the computed descriptors. To compute molecular descriptors, we'll use the dataset from the previous stage and the SMILES notation (which represents the unique chemical structure of molecules). Lipinski's descriptors are the descriptors that we shall compute (molecular weight, LogP, number of hydrogen bond acceptors and number of hydrogen bond donors). Finally, we'll use box plots to undertake exploratory data analysis to see how the active and inactive sets of substances differ. We are selecting only two bioactivity classes which are active or inactive, so the intermediate class is removed from the data set. The Lipinski descriptors are a set of rules which are used to evaluate the drug likeliness of the compounds. The drug's likeliness is based on pharmacokinetic properties such as Absorption, Distribution, Metabolism and Excretion (ADME properties). These properties will represent the drug likeliness of a particular compound whether it could be absorbed into the body, distributed to the tissue and organs and become metabolized and eventually excreted from the body. Lipinski's Rule stated the following: Molecular weight <500 Dalton; LogP <5; Hydrogen bond donors <5; Hydrogen bond acceptors <10.

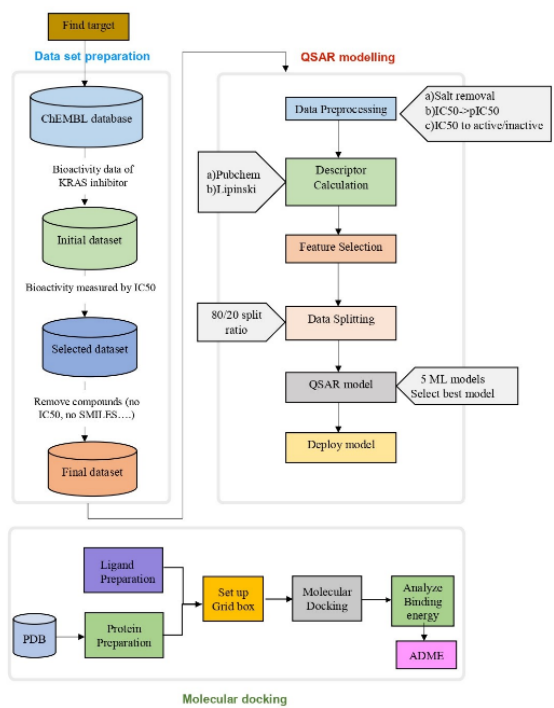

**Figure 3.** Workflow of model creation and molecular docking.

Python's kit package is used to calculate the descriptors. SMILES notation is input to the function to calculate Lipinski descriptors. Next, convert the IC50 value to the pIC50 value.

$$IC50 = -log10(IC50)$$

The IC50 values with large values will become negative values after performing a negative logarithm. So, the maximum value is set as 100,000,000 otherwise the negative logarithmic value will become negative. So that no IC50 value will be greater than 100,000,000. After performing a negative logarithmic transformation no value will be negative. All the pIC50 values will be greater than 1.0.

EDA is performed using Lipinski descriptors. EDA is a chemical space analysis because it allows us to look at the chemical space. The frequency plot of two bioactivity classes is shown in Figure 4. The boxplot for the pIC50 value is shown in Figure 5. We use a threshold to define active and inactive. The threshold used is 5 for inactive and 6 for active. If pIC50 value is greater than 6 it will be active. If it is less than 6 it will be inactive.

### 3.2 Mann-Whitney U Test

Mann-Whitney U test is performed to look at the difference between two bioactivity classes: active and inactive. Mann-Whitney U test is used to test the statistical significance of the difference whether they are different or not different. Mann-Whitney function is applied to pIC50 values, then it will compare the active class and the inactive class whether there is a statistical significance for the pIC50 variable. Based on this analysis if the p-value is low then we reject the null hypothesis. So, we can say that it is having different distribution. We are performing the Mann-Whitney U test for all four Lipinski descriptors also. The graphical representation of the results of the Mann-Whitney U test is given in Figure 6.

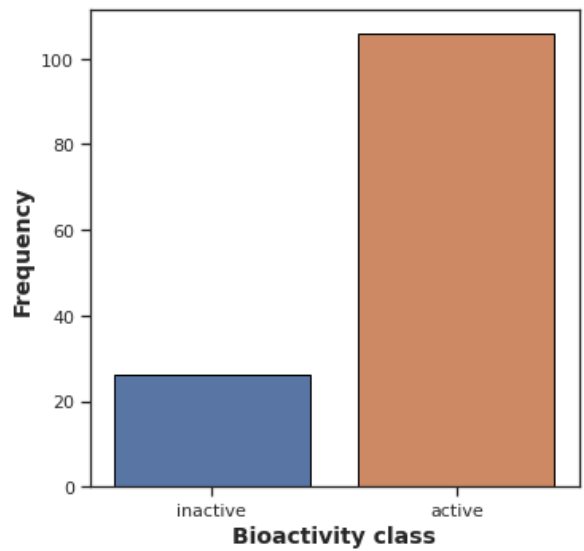

**Figure 4.**   Frequency plot of active and inactive classes.

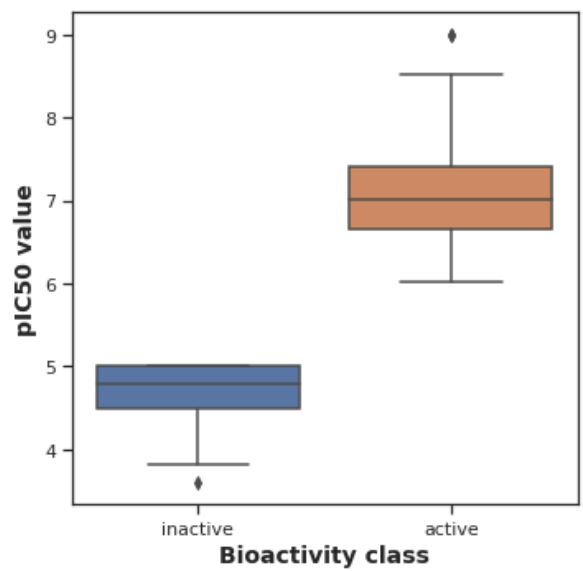

**Figure 5.**   Box plot of bioactivity class and pIC50 value.

Taking a look at pIC50 values, the actives and inactive displayed statistically significant differences, which is to be expected since threshold values (IC50 <1,000 nM = Actives while IC50 >10,000 nM = Inactives, corresponding to pIC50 >6 = Actives and pIC50 <5 = Inactives) were used to define actives and inactives. Of the four Lipinski's descriptors (MW, LogP, NumHDonors and NumHAcceptors), only MW, NumHDonors and NumHAcceptors exhibited no difference (same distribution) between the actives and inactives while the LogP descriptor shows a statistical significant difference between actives and inactives.

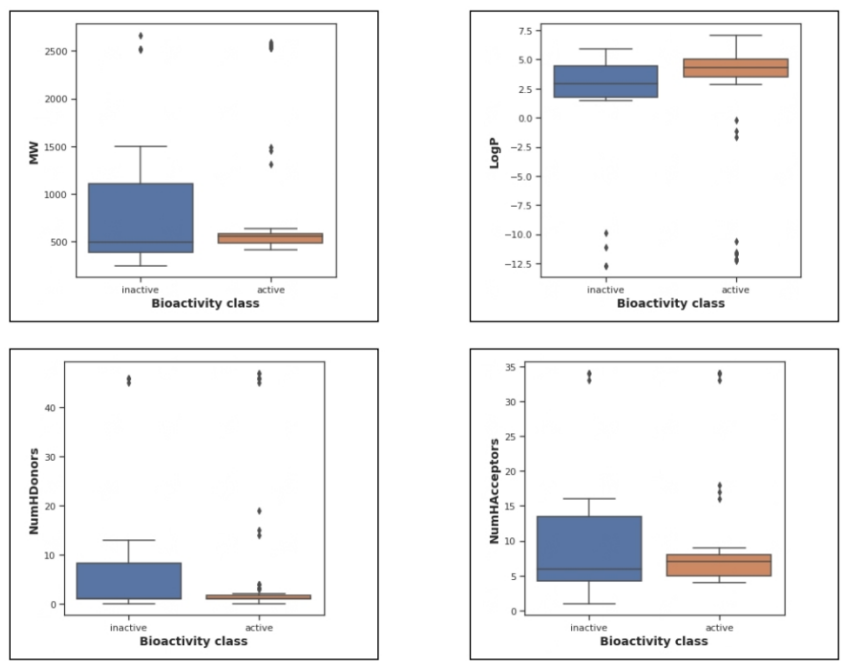

**Figure 6.** Boxplot of each Lipinski descriptors against bioactivity class.

### 3.3 Fingerprint Calculation

The molecular descriptors are computed using the PADEL-Descriptor software. Canonical SMILES of the compounds are used for this. The software performs salt removal and after cleaning the structure it will perform descriptor calculation.

The selected fingerprint type is PubChem fingerprint. The Lipinski descriptors will describe the global features of the molecule and the PubChem fingerprints will describe the local features of the molecule. A fingerprint is an ordered list of binaries (1/0) bits. The PubChem fingerprint encodes molecular fragment information with 881 binary digits. These fingerprints will help us to uniquely identify each molecule. PubChem fingerprints and pIC50 values are used for model building.

### 3.4 Model Building

We use the computed molecular descriptors to build regression model for predicting the pIC50 values. PubChem fingerprints have 881 input features. Each molecule has unique properties. Each molecule will allow the machine learning algorithm to learn from the unique properties and then create a model that can distinguish between compounds that are active or inactive. We want to see which functional groups or fingerprints are essential for designing a good drug. The target variable used for prediction is pIC50. After the removal of low variance features, 137 fingerprints are left from the 881 fingerprints.

The data is split in an 80/20 ratio. Then five machine learning models are created. They are Random Forest Regressor, SVR Regression, Ridge Regression, Decision Tree Regression, and Linear Regression. Among these, the best model is identified for bioactivity prediction.

Model validation is an important process, which should be performed to ensure that a fitted model can accurately predict responses for future or unknown subjects. Two statistical parameters were used to evaluate the performance of the QSAR models consisting of Root Mean Squared Error (RMSE) and Coefficient of determination ($r^2$). The r value is a commonly used metric to represent the degree of relationship between two variables of interest. The best model is deployed for predicting the bioactivity of different compounds. Using this model, we can analyse the bioactivity data of selected six ligands.

### 3.5 Creation of Web Application

After finding out the best regression model based on RMSE and $r^2$ values the web application is created. The web application is created using Python's Streamlit library. This web application will predict the bioactivity (pIC50 values) of each compound. A text file which consists of CHEMBL id and SMILES notation of compounds is given as input and the application will output the corresponding bioactivity.

### 3.6 Bioactivity Prediction

From fifty ligands after molecular docking, we have selected six ligands which have the least binding energy value. They are quinic acid, beta citronellol, Gallic acid, borneol, esculetin and geraniol. Further studies are based on these six ligands. From creating five machine learning models based on RMSE and $r^2$ values, the Random Forest regression model is the best one. So, we can use this model for bioactivity prediction. A comparison of RMSE and $r^2$ is given in Table 3. Based on this the best models are which having low values for RMSE and high values for $r^2$. From Table 3 Random Forest Regressor and SVR Regression models are best. But SVR is not suitable for large datasets. So, we select Random Forest Regressor for building web applications. The interface of the web application is shown in Figure 7. The web application is created using the Random Forest Regression model. The bioactivity of selected ligands is predicted using the created web application. One with the highest value is the best. The predicted bioactivity of ligands is given in Table 2.

**Figure 7.** Interface of the web application.

Table 2.  Bioactivity data of selected six ligands

| Molecule name | pIC50 |
|---|---|
| CHEMBL465398 | 5.123973 |
| CHEMBL395827 | 5.049509 |
| CHEMBL288114 | 5.072908 |
| CHEMBL486208 | 5.295054 |
| CHEMBL244743 | 5.0961963 |
| CHEMBL25719 | 4.942212 |

Table 3.  RMSE and $r^2$ values of five models

| Model name | RMSE | $R^2$ |
|---|---|---|
| Random Forest Regressor | 0.955 | 0.1589 |
| SVR Regression | 1.022 | 0.3921 |
| Ridge Regression | 0.884 | 0.259 |
| Decision Tree Regression | 1.007 | -0.089 |
| Linear Regression | 49.106 | -1.870e+23 |

We have given the 6 selected ligands together with their molecular ChEMBL id as an input to the bioactivity prediction web app. The Simplified Molecular-Input Line-Entry System (SMILES) notation and ChEMBL id of six ligands is input to the web application. The model predicts the pIC50 value of each of the ligands. Due to the minus sign (-log IC50), higher values of pIC50 indicate exponentially more potent inhibitors. Here we found that borneol is having the highest pIC50 value among others. So, it is a more potent inhibitor.

## 3.7  ADME Analysis

ADME properties and drug likeliness of selected ligands are studied using the SwissADME tool. The Blood-Brain Barrier (BBB), pharmacokinetics, drug-likeness and medicinal chemistry friendliness of compounds are evaluated additionally. After studying Lipinski values, a table is made based on that which is shown in Table 4. The table describes the name of the ligand and binding energies together with the Lipinski descriptors, physicochemical properties, Pharmacokinetics, Drug likeness, and Medicinal Chemistry of six ligands. The druglike classifiers defined by Lipinski[22], Veber[23], Ghose[24], Egan[25], and Muegge[26] are used here. Chemical substances known as Pan-Assay Interference Compounds (PAINS) frequently produce false positive results[27]. PAINS have a tendency to respond non-specifically with a variety of biological targets rather than impacting a single target. Toxoflavin, isothiazolinones, quinones, and catechols are some of the most common PAINS[27]. The Blood-Brain Barrier (BBB), which is made up of endothelial cells, is a highly selective semipermeable border that prevents the non-selective movement of solutes from the bloodstream into the extracellular fluid of the central nervous system, where neurons are found. A microvascular structure called the BBB controls drug permeability to the brain in a targeted manner. P-gp (P-glycoprotein 1) is highly expressed on the surface of the cancer cells and acts as an efflux pump, preventing drug accumulation inside the tumour. Anticancer medicines are ejected before they reach their desired target. It also plays a role in the development of anticancer drug resistance in cells. The rate at which a substance penetrates the stratum corneum is measured by skin permeability (Kp). This number is commonly used to quantify the movement of molecules in the epidermal skin's outermost layer and to highlight the importance of skin absorption. The Topological Polar Surface Area (TPSA) of a molecule is defined as the surface sum over all polar atoms or molecules, primarily oxygen and nitrogen, including their connected hydrogen atoms.

The drug like classifiers such as Lipinski[22], Veber[23], Ghose[24], Egan[25] and Muegge[26] are analysed. Ghose suggested a qualifying range that could be used for the development of drug like chemical libraries and he has recommended the following constraints: Molecular Weight (MW) between 160 and 480; calculated logP between -0.4 and 5.6; molar refractivity between 40 and 130 and the total number of atoms between 20 and 70. Veber's Rule has a 500 molecular weight cut-off. The compounds which meet only the two criteria of 10 or fewer rotatable bonds and a polar surface area no greater than 140 Å$^2$ are predicted to have good oral bioavailability. The Egan rule considers good bioavailability for compounds with $0 \geq tPSA \leq 132$ Å$^2$; $-1 \geq WlogP \leq 6$. Muegge's rule considers the criteria such as Molecular weight between 200 and 600; LogP between 2 and 5; TSPA ≤ 150; the number of rings ≤ 7; number of carbons>4; number of heteroatoms>1; number of Rotatable Bonds (ROTB) ≤ 15; Hydrogen Bond Donors (HBD) ≤ 5 and hydrogen bond acceptors (HBA) ≤ 10. For defining lead-like compounds, the Rule of Five (RO5) has been extended to the Rule of Three (RO3)[28]. Rule of three is defined as following characteristics: log P should not be greater than 3 molecular mass less than 300 Daltons; HBD not more than 3; HBA not more than 3 and ROTB not more than 3. The table which represents these properties is shown in Table 4. From Table 4 all the six ligands follow the Lipinski rule of five. A low LogP value of quinic acid indicates good permeation and absorption with higher hydrophilicity. But it is not BBB permeant. Beta citronellol and borneol are BBB permeant.

Beta citronellol, borneol, geraniol are BBB permeant. Quinic acid is a P-gp substrate and others are non-P-gp substrates. SwissADME analysis revealed that all of the ligands adhered to Veber and Egan's drug-like filters, which defined drug-likeness constraints using several criteria. Beta citronellol, borneol, esculetin, and Geraniol each obtained a 0.55 bioavailability score, indicating a chance of 55% (greater than 10%) for rat bioavailability and bioavailability of quinic acid and gallic acid is 0.56. The bioavailability score predicts the probability of a compound having at least 10% oral bioavailability in the rat. The majority of the substances have bioavailability scores of 0.55 or 0.56, indicating that they have favourable pharmacokinetic characteristics. Supplements with high bioavailability will be more effective since they will help the body absorb more of the necessary vitamin without requiring larger doses. Quinic acid and borneol did not produce any PAINS or Brenk alerts, demonstrating the specificity of each compound. In comparison to the other ligands under study, gallic acid showed a lower value of synthetic accessibility.

**Table 4.** Physicochemical properties, drug likeliness and pharmacokinetics prediction of all the six ligands

|  | Quinic acid | Beta citronellol | Gallicacid | Borneol | Esculetin | Geraniol |
|---|---|---|---|---|---|---|
| **Binding Energy** | +0.70 | +1.71 | +2.63 | +3.51 | +4.53 | +5.83 |
|  |  |  |  |  |  |  |
| **Physicochemical Properties** |  |  |  |  |  |  |
| Formula | $C_7H_{12}O_6$ | $C_{10}H_{20}O$ | $C_7H_6O_5$ | $C_{10}H_{18}O$ | $C_9H_6O_4$ | $C_{10}H_{18}O$ |
| H-bond donors | 5 | 1 | 4 | 1 | 2 | 1 |
| H-bond acceptors | 5 | 1 | 5 | 1 | 4 | 1 |
| Heavy atoms | 13 | 11 | 12 | 0 | 13 | 0 |
| rotatable bonds | 1 | 5 | 1 | 0 | 0 | 4 |
| Molar refractivity | 40.11 | 50.87 | 39.47 | 46.60 | 46.53 | 50.40 |
| TPSA | 118.22 ˚$A^2$ | 20.23 ˚$A^2$ | 97.99 ˚$A^2$ | 20.23 ˚$A^2$ | 70.67 ˚$A^2$ | 20.23 ˚$A^2$ |
|  |  |  |  |  |  |  |
| **Pharmacokinetics** |  |  |  |  |  |  |
| GI absorption | Low | High | High | High | High | High |
| BBB permeant | No | Yes | No | Yes | No | Yes |
| P-gp substrate | Yes | No | No | No | No | No |
|  |  |  |  |  |  |  |
| **Druglikeness** |  |  |  |  |  |  |
| Lipinski | Yes | Yes | Yes | Yes | Yes | Yes |
| Ghose | No | No | No | No | No | No |
| Veber | Yes | Yes | Yes | Yes | Yes | Yes |
| Egan | Yes | Yes | Yes | Yes | Yes | Yes |
| Muegge | No | No | No | No | No | No |
| Bioavailability score | 0.56 | 0.55 | 0.56 | 0.55 | 0.55 | 0.55 |
|  |  |  |  |  |  |  |
| **Medicinal Chemistry** |  |  |  |  |  |  |
| PAINS | False | False | True | False(0 alert) | True | False |
| Brenk | False | True | True | False | True | True |
| Synthetic accessibility | 3.34 | 2.61 | 1.22 | 3.43 | 2.61 | 2.58 |

Figure 8 depicts the bioavailability radar for all selected six ligands, which includes factors like size, flexibility, insolubility, lipophilicity, unsaturation and polarity. The coloured region represents the relevant physicochemical space for oral bioavailability. The bioavailability radar enables a first glance at the drug-likeness of a molecule. The pink area represents the optimal range for each property (lipophilicity: XLOGP3 between 0.7 and +5.0, size: MW between 150 and 500 g/mol, polarity: TPSA between 20 and 130 A2, solubility: log S not higher than 6, saturation: fraction of carbons in the sp3 hybridization not less than 0.25, and flexibility: no more than 9 rotatable bonds. Because of the significant unsaturation, it is projected that esculetin and gallic acid will not be orally available in this case. The bioavailability radar of quinic acid, beta citronellol, borneol, and geraniol is fully inside and that is they have optimized properties for bioavailability.

From Figure 9 the BOILED-Egg, is a method for predicting two key ADME parameters at the same time, namely passive gastrointestinal absorption (HIA) and Blood-Brain Barrier (BBB). Although it is conceptually simple because it just uses WLOGP and TPSA (two physicochemical descriptors), for lipophilicity and apparent polarity). By color-coding the graph, the user can obtain a global assessment of passive brain access (inside/outside the egg yolk), passive absorption (inside/outside the egg white), and active efflux from the Central Nervous System (CNS) or the gastrointestinal lumen. The blue dots indicate P-gp substrates (PGP+), and the red dots indicate P-gp non-substrates (PGP-). The yellow portion (yolk) is for a high probability of brain penetration, and the white region is for high probability of passive absorption by the gastrointestinal tract. Yolk and white areas are not mutually exclusive. The points are coloured blue if the compound is predicted as actively effluxed by P-gp (PGP+) and red if predicted as non-substrate of P-gp (PGP-). The points outside the egg are predicted as non-absorbent or non-penetrant (outside egg). Here quinic acid is predicted as not absorbed and not brain penetrant (outside the Egg), borneol, beta citronellol, and geraniol is predicted as brain-penetrant (inside the egg yolk) and it is not subject to active efflux, P-gp- (red dot).

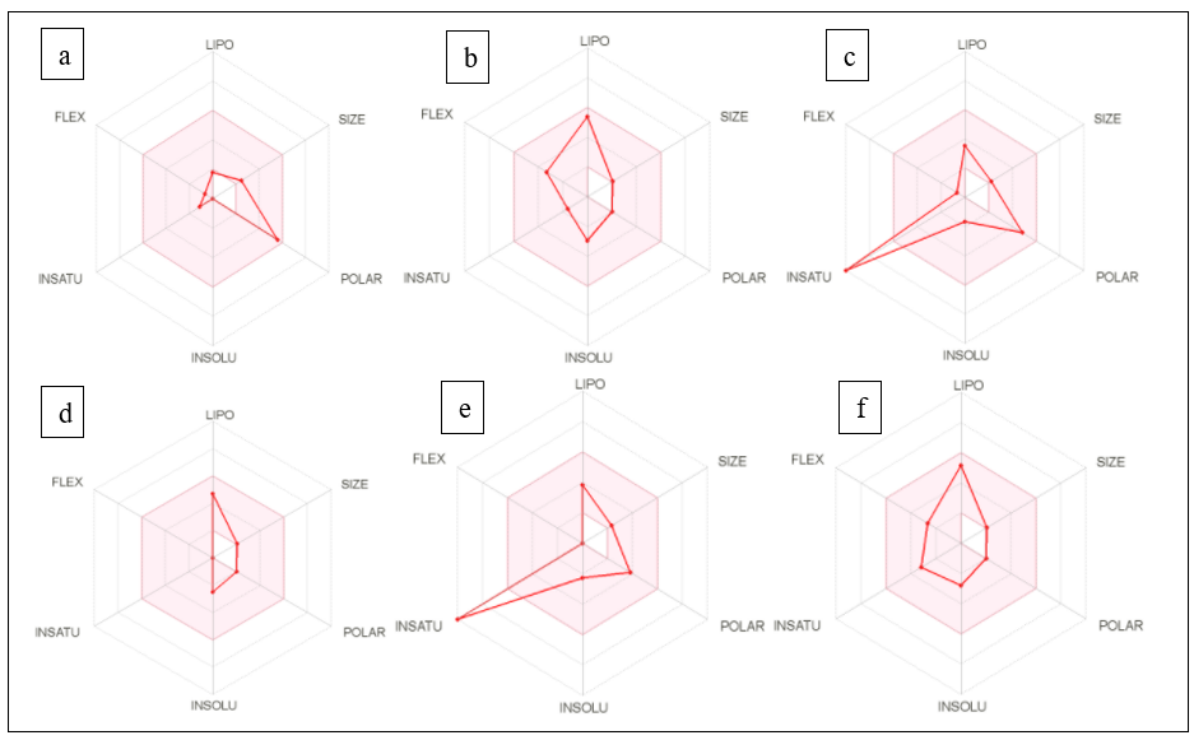

**Figure 8.**   The bioavailability radar of six ligands, analysed using SwissADME web tool.

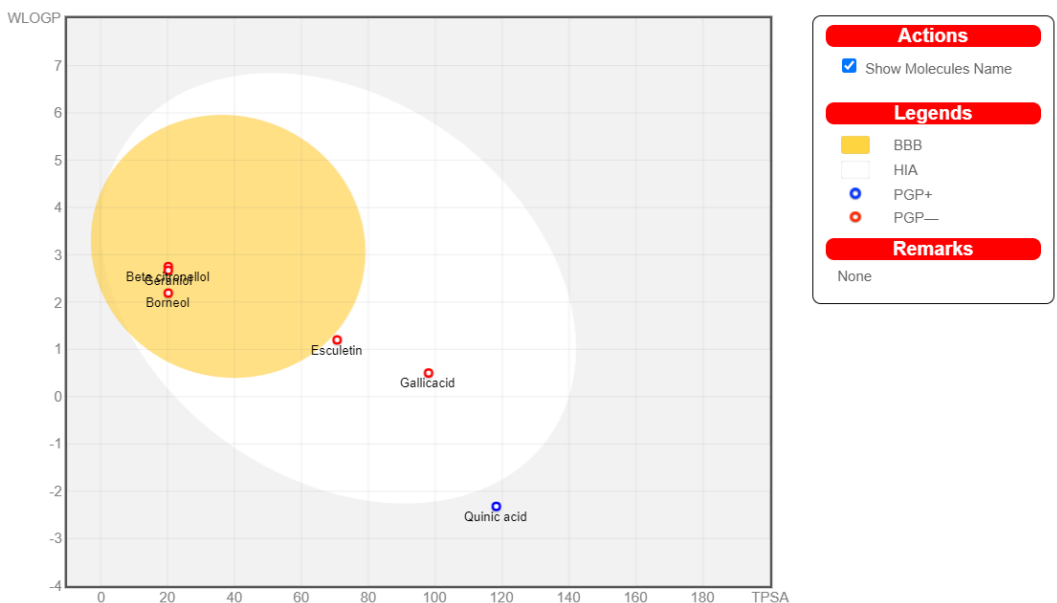

**Figure 9.**  The BOILED-Egg model is used for predicting HIA and BBB.

Quinic acid is predicted to not be absorbed and not enter the brain (outside egg), while gallic acid and esculetin are predicted to be effectively absorbed but not to reach the brain (in egg white) and it is PGP- (red dot). Here quinic acid is PGP+ that is they will be ejected before they reach their desired target. So, absorption is not sure for quinic acid. gallic acid and esculetin are non-BBB permeants and non-PGP substrates. But they have high insaturation from Figure 8 bioavailability radar. The remaining ligands are beta citronellol, borneol, and geraniol. Beta citronellol and geraniol were visualized as one alert for Brenk. borneol was visualized as no alert for both PAINS and Brenk. Based on all these factors we analysed borneol as the best ligand.

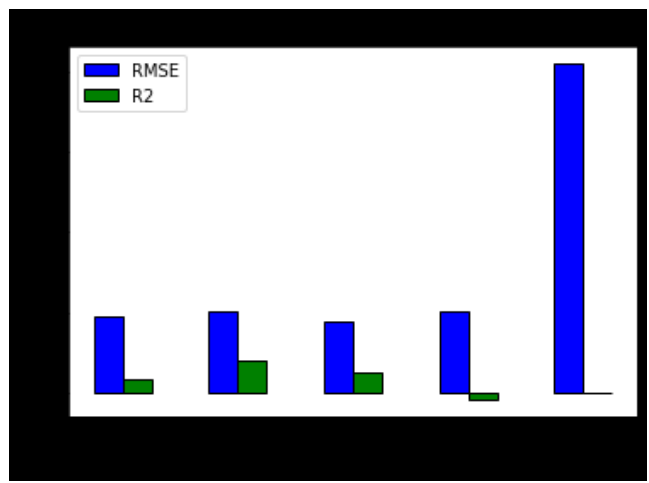

**Figure 10.**  Comparison of five regression models.

## 4. Performance Analysis

Five machine learning models are used for creating the bioactivity prediction application. They are Random Forest regression, SVM, Ridge Regression model, Decision Tree regression model and Linear Regression model. Each model's performance is analysed based on the RMSE and $r^2$ values.

The one with least RMSE value and the highest value for $r^2$ is the best model. From Figure 10 and Table 3

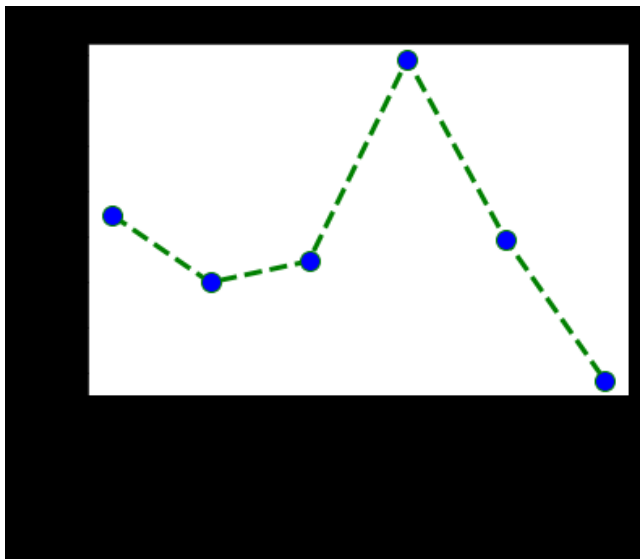

**Figure 11.** Comparison of six ligands based on bioactivity.

Random Forest regressor is the best machine learning selected for our work.

The graphical plot of the bio-activity of each of the six ligands is given in Figure 11. Bioactivity (pIC50) value should be high. The ligand with high value is best. From graph Figure 11 borneol is having the highest value for Bioactivity (pIC50) value. So, it has the best bioactivity score.

## 5. Discussion

In this study, the compound borneol found in ginger, sage, rosemary, and mint is proficient in binding activity to KRAS (PDB ID: 6MNX). Many plants present in nature have anti-cancer properties. Synthetic drugs have more side effects. Drugs approved by the FDA for the treatment of Pancreatic Cancer are Gemcitabine (Gemzar), 5-fluorouracil (5-FU) or Capecitabine (Xeloda) (an oral 5FU drug), Irinotecan (Camptosar) etc. These drugs can be used for Chemotherapies. But these have huge side effects such as vomiting, nausea, hair loss, loss of appetite, etc. Drugs such as Paclitaxel, Oxaliplatin, etc. used for chemotherapy can even damage the nerves. Natural plants can be used as drugs. It has fewer side effects. Here selected ligands were fifty compounds found in different plants. After initial screening, we selected six ligands. These ligands have anti-cancer properties and showed better binding results with KRAS. However, borneol performed best in docking and also drug likeliness analysis. It also showed good results in ADME analysis compared to other ligands. Five machine learning models were created to analyse the bioactivity of any compounds when they interact with the target protein KRAS. From these models, Random Forest is chosen as the best model based on RMSE and $r^2$ values. The bioactivity prediction application is built using the Random Forest regression model as the backbone. Borneol can achieve inhibition of KRAS which will further control the uncontrollable cell growth. Further experiments and studies may be performed to strengthen the studies' results. This study aims to find out plant-derived compounds that can be used for treating pancreatic cancer using efficient computer aided methods and Machine Learning.

## 6. Conclusion

Despite many efforts, the development of effective drugs for cancer will take considerable time. Machine Learning (ML) offers promising solutios that could accelerate the discovery and optimiztion of new drugs. The study describes the development of the machine learning model for users to compute the bioactivity of a diverse range of compounds having inhibitory effects against KRAS. The study facilitates the prediction of bioactivity values for compounds based on the QSAR models without performing any biological assays. The prediction method is based on the molecular descriptors generated from the QSAR equation. The Random Forest regression model will serve as a useful screening method for pharmacologists and medicinal chemistry to screen any novel compounds thought to have a KRAS inhibitory activity before jumping into *in vitro* experimental assays. This technique is expected to be used further in drug design and development strategies. To gain a further understanding of the interaction between KRAS and its inhibitors, a chemically diverse set of some representative compounds can be extracted from active KRAS inhibitors (i.e., having IC50 <1 uM) using the Kennard Stone algorithm and subjected to an investigation on its binding modality against the active site of KRAS using molecular docking. Here we have selected fifty plant-derived compounds from natural plants which are easily available in our environment and performed docking studies and ADME studies. By considering all the factors we found borneol as the best ligand among all the ligands and borneol is found in natural plants like ginger, sage, rosemary, and mint. Among the tested ligands, borneol holds promise as an efficient chemotherapeutic agent.